\documentclass{PoS}

\title{Strong coupling constant from moments of quarkonium correlators}

\ShortTitle{Strong coupling constant}

\author{\speaker{Peter Petreczky}\thanks{This work was supported by U.S. Department of Energy under 
Contract No. DE-SC0012704. I would like to thank Yu Maezawa and Johannes Weber the collaboration on
        this topic.}\\
        Physics Department, Brookhaven National Laboratory, Upton, NY 11973, USA\\
        E-mail: \email{petreczk@bnl.gov}}

\abstract{I review the determination of the strong coupling constant from moments
          of quarkonium correlators calculated on the lattice. I discuss different
          sources of systematic errors in such calculations.
          }

\FullConference{XIII Quark Confinement and the Hadron Spectrum - Confinement2018\\
		31 July - 6 August 2018\\
		Maynooth University, Ireland}

\begin{document}

\section{Introduction}
The strong coupling constant is a fundamental parameter of QCD and its knowledge is
needed to make predictions in the Standard Model. The Particle Data Group gives
the value $\alpha_s(M_Z,n_f=5)=0.1182(12)$ \cite{PDG18}, which has a small error, but the scattering of
individual determinations around the central value is much larger than the quoted
error. This could possibly mean that the errors in the determination of $\alpha_s$
are not completely under control. Lattice QCD calculations may help to obtain
an accurate value for $\alpha_s$ as these calculations are becoming more and more precise.
The comparison of the strong coupling constant obtained from lattice and non-lattice
methods is important not only for understanding the systematic errors in $\alpha_s$ determination but also for 
establishing the connection between Euclidean lattice QCD and perturbative QCD in the time-like
region.

There are several quantities calculable on the lattice that are suitable for
extracting the strong coupling constant. These include the 
small Wilson loops \cite{Davies:2008sw,Maltman:2008bx,McNeile:2010ji}, 
moments of quarkonium correlators \cite{McNeile:2010ji,Allison:2008xk,Chakraborty:2014aca,Maezawa:2016vgv,Nakayama:2016atf}, 
hadronic vacuum polarization \cite{Shintani:2008ga,Shintani:2010ph,Hudspith:2015xoa,Hudspith:2018bpz}, the static
quark anti-quark energy \cite{Bazavov:2012ka,Bazavov:2014soa,Takaura:2018vcy,Takaura:2018lpw} and the Schr\"odinger functional \cite{Bruno:2017gxd}. 
Furthermore, there have been attempts to extract $\alpha_s$ using eigenvalue spectrum of Dirac the operator \cite{Nakayama:2018ubk}
and the quark-gluon and the gluon-ghost
vertices from calculations in fixed gauge \cite{Blossier:2013ioa,Sternbeck:2012qs}. Two of the methods
to determine $\alpha_s$ from the lattice, namely the Schr\"odinger functional approach
and the calculation of the static quark anti-quark energy are discussed in a different
contribution to these proceedings \cite{Pich:2018lmu}. Here I will focus on the determination of the strong
coupling constant from the moments of quarkonium correlators. In fact the determination of $\alpha_s$ from
the moments of quarkonium correlators predates lattice QCD calculations (see e.g. \cite{Kuhn:2001dm}) since
moments of quarkonium correlators in the vector channel can be extracted from the experimental data on $e^{+} e^{-}$
collisions. Early lattice studies of the moments of charmonium correlators with the aim of obtaining the charm
quark mass have been reported in Refs. \cite{Bochkarev:1995ai,Bochkarev:1996sg}.

\section{Moments of quarkonium correlators and the strong coupling constant}

One can consider moments of quarkonium correlators in different channels, e.g.
vector, pseudo-scalar, scalar etc. Since the pseudo-scalar correlators are the least
noisy in the lattice calculations it makes sense to consider the moments of pseudo-scalar correlators for
the precision determination of the strong coupling constant.
The moments of the pseudo-scalar quarkonium correlator,
are defined as
\begin{equation}
G_n=\sum_t t^n G(t),~G(t)=a^6 \sum_{\mathbf{x}} (a m_{h0})^2 \langle j_5(\mathbf{x},t) j_5(0,0) \rangle.
\end{equation}
Here $j_5=\bar \psi \gamma_5 \psi $ is the pseudo-scalar current, $a$ is the lattice spacing and $m_{h0}$ is the bare lattice heavy quark mass.
On the lattice the above definition is modified 
in the following way:
\begin{equation}
G_n=\sum_t t^n (G(t)+G(N_t-t)).
\label{eq:latmom}
\end{equation}
The moments $G_n$ are finite only for $n \ge 4$ ($n$ even) in the $a\rightarrow 0$ limit since the correlation function 
diverges as $t^{-4}$ for small $t$. Furthermore, the moments $G_n$ do not
need renormalization because
 the explicit factors of the quark mass are included in
their definition \cite{Allison:2008xk}.
The moments can be calculated in perturbation theory in $\overline{MS}$ scheme
\begin{equation}
G_n=\frac{g_n(\alpha_s(\mu),\mu/m_h)}{a m_h^{n-4}(\mu_m)}.
\end{equation}
Here $\mu$ is the $\overline{MS}$ renormalization scale, $m_h(\mu_m)$ is the renormalized
heavy quark mass in $\overline{MS}$ scheme. The scale $\mu_m$ at which the $\overline{MS}$ 
heavy quark mass is defined can be different from $\mu$ \cite{Dehnadi:2015fra}, though most studies
assume $\mu_m=\mu$.
The coefficient $g_n(\alpha_s(\mu),\mu/m_h)$ is calculated up to 4-loop, i.e. up to order $\alpha_s^3$
\cite{Sturm:2008eb,Kiyo:2009gb,Maier:2009fz}.
Given the lattice data on $G_n$ one can extract $\alpha_s(\mu)$ and $m_h(\mu)$ from the above
equation. However, as discussed in Ref.~\cite{Allison:2008xk} it is more practical to
consider the reduced moments
\begin{equation}
R_n =\left\{ \begin{array}{ll}
G_n/G_n^{(0)} & (n=4) \\
\left(G_n/G_n^{(0)}\right)^{1/(n-4)} & (n\ge6) \\
\end{array} \right.
\label{eq:redmom},
\end{equation}
where $G_n^{(0)}$ is the moment calculated from the free lattice correlation function.
The lattice artifacts largely cancel out in the reduced moments.

It is straightforward to write down the perturbative expansion for $R_n$:
\begin{eqnarray}
R_n &=& \left\{ \begin{array}{ll}
r_4 & (n=4) \\
r_n \cdot \left({m_{h0}}/{m_h(\mu)}\right) & (n\ge6)\\
\end{array}\right. , \label{rn_pert}\\
r_n &=& 1 + \sum_{j=1}^3 r_{nj}(\mu/m_h) \left(\frac{\alpha_s(\mu)}{\pi}\right)^j.
\end{eqnarray}
There is also a contribution to the moments of quarkonium
correlators from the gluon condensate \cite{Broadhurst:1994qj}.
From the above equations it is clear that $R_4$ as well as the ratios $R_6/R_8$ and $R_8/R_{10}$
are suitable for the extraction of the strong coupling constant $\alpha_s(\mu)$, while
the ratios $R_n/m_{h0}$ with $n\ge 6$ are suitable for extracting the heavy quark mass $m_h(\mu)$.

There are several calculations of the moments of quarkonium correlators. The first such
calculation was performed by HPQCD Collaboration using asqtad improved staggered action for two flavors
of light quarks and a strange quark in the sea,
and Highly Improved Staggered Quark (HISQ) action for the heavy valence quark \cite{Allison:2008xk}.
Here the heavy quark mass was equal to the charm quark mass. Later this calculation was extended using
smaller lattice spacing and more values of the valence heavy quark mass \cite{McNeile:2010ji}. The most
recent calculation by HPQCD used 4 flavor lattice simulations, i.e. simulations including the effect
of dynamical charm quark with HISQ action in the sea and valence sectors and several heavy 
valence quark masses \cite{Chakraborty:2014aca}.
There is also a calculations of the moments of the charmonium correlators using 3 flavors of HISQ
sea quarks (two light quarks and a strange quark) and HISQ valence charm quark \cite{Maezawa:2016vgv}. The main feature of
this study is that many lattice spacings have been used. Very recently the 3 flavor HISQ calculation was
extended to include finer lattices and several values of the heavy quark mass larger than the charm
quark, namely $m_h=m_c,~1.5m_c,~2m_c$ and $3m_c$, $m_c$ being the charm quark mass \cite{Petreczky:2019ozv}. 
Finally there is also a calculation of the moment of charmonium correlators 
using domain wall fermions in the 3 flavor sea as well as in the valence sector \cite{Nakayama:2016atf} at three values of the lattice spacings.

\begin{figure}
\includegraphics[width=7.5cm]{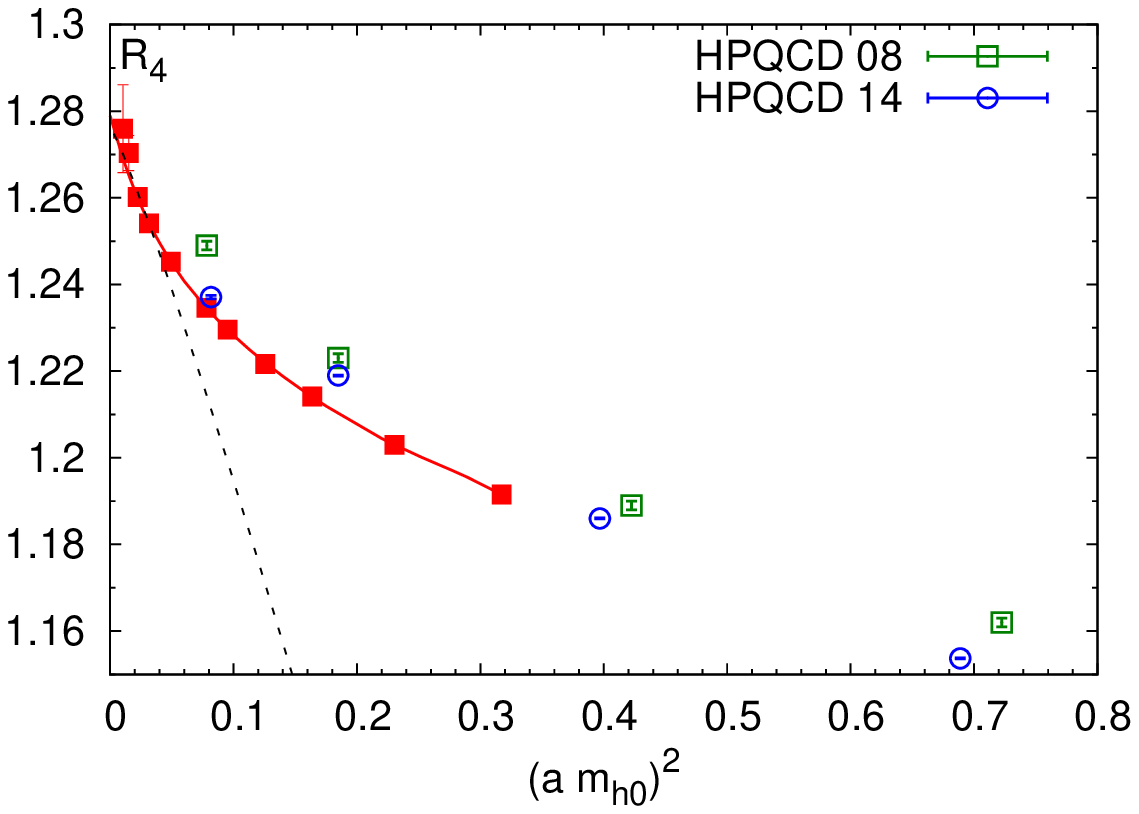}
\includegraphics[width=7.5cm]{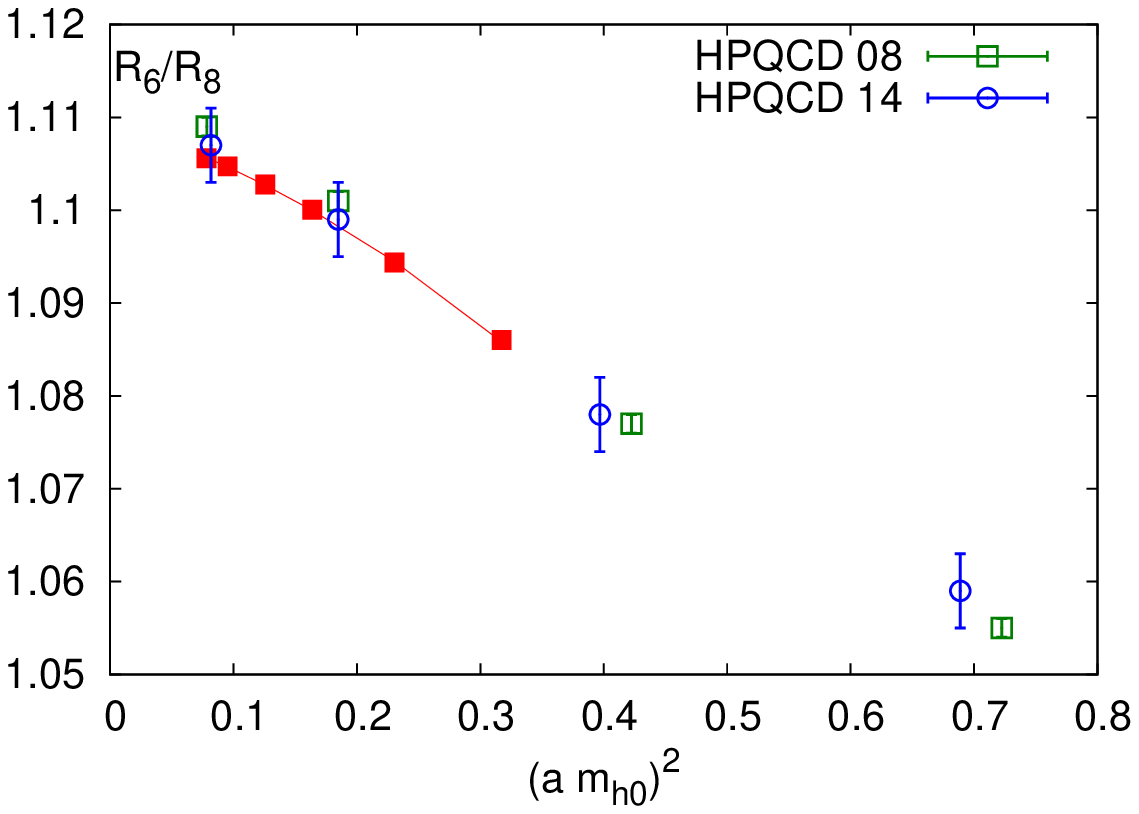}
\caption{The lattice spacing dependence of $R_4$ and $R_6/R_8$ for $m_h=m_c$. The filled symbols
correspond to the lattice results of Ref. \cite{Petreczky:2019ozv}, while the open symbols correspond to HPQCD results \cite{Allison:2008xk,Chakraborty:2014aca}.
The solid line corresponds to polynomial fit, see text. The dashed line corresponds to simple $a^2$ fit.
The errors for HPQCD 14 result for $R_6/R_8$ have been obtained by propagating the errors on $R_6$ and $R_8$.}
\label{fig:adep}
\end{figure}

One of the challenge for accurate determination of the strong coupling constant from the moments
of quarkonium correlators is to obtain reliable continuum extrapolations for the reduced moments. I will demonstrate this problem
using the very recent results from Ref. \cite{Petreczky:2019ozv} as an example.
The lattice spacing dependence of $R_4$ and $R_6/R_8$ is shown in Fig. \ref{fig:adep} for $m_h=m_c$.
One can see that there is a significant dependence on $a^2$ which cannot be described by a simple
$a^2$ form. Since the tree level lattice artifacts are canceled out in the reduced moments the discretization
errors should scale like $\alpha_s^n (a m_{h0})^{2j}$. In order to fit the lattice spacing dependence
of $R_4$ of all available lattice data  one should consider terms up to fifth order 
in $(a m_{h0})^2$ and second order in $\alpha_s$ \cite{Petreczky:2019ozv}. For the ratio $R_6/R_8$ lower order
polynomials can be used. 
For the four smallest lattice spacings a simple $a^2$ extrapolation can be used for $R_4$,
which agrees with the above extrapolations within the errors, see Fig. \ref{fig:adep}.
Many different continuum extrapolations have been performed in Ref. \cite{Petreczky:2019ozv} and the
differences in the corresponding continuum results have been used as estimates of systematic errors.
In Fig. \ref{fig:adep} I also show the HPQCD results for the reduced moments
for heavy quark mass around the charm quark mass. Here far fewer lattice data are available 
and Bayesian fits had to be used to perform the continuum extrapolation \cite{Allison:2008xk,McNeile:2010ji,Chakraborty:2014aca}.

\begin{figure}
\includegraphics[width=4.7cm]{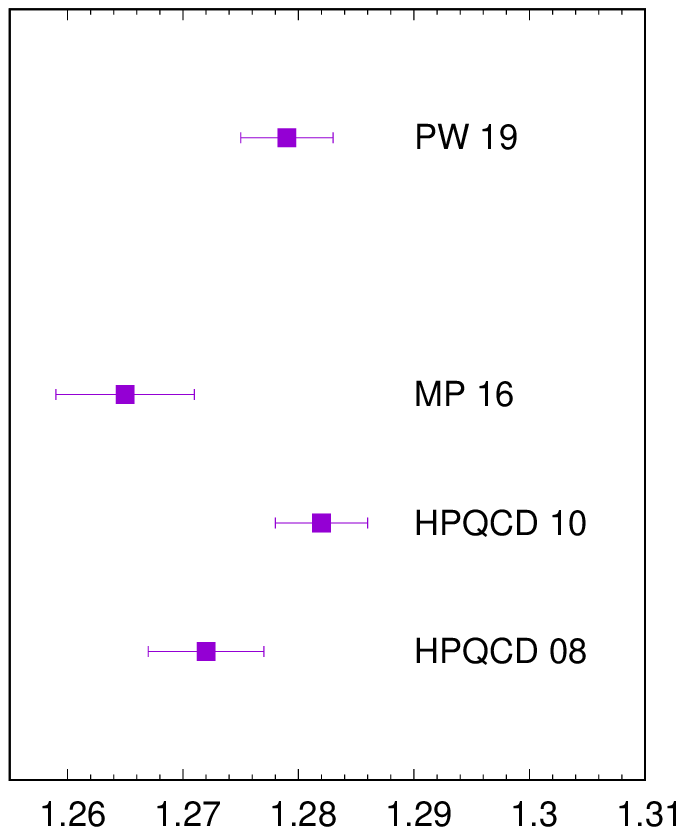}
\includegraphics[width=4.7cm]{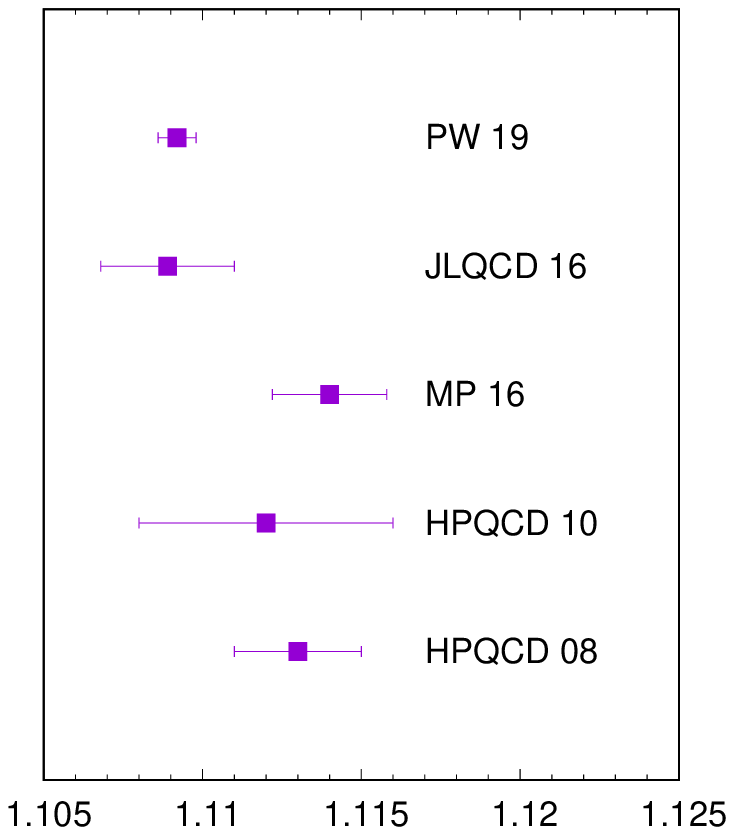}
\includegraphics[width=4.7cm]{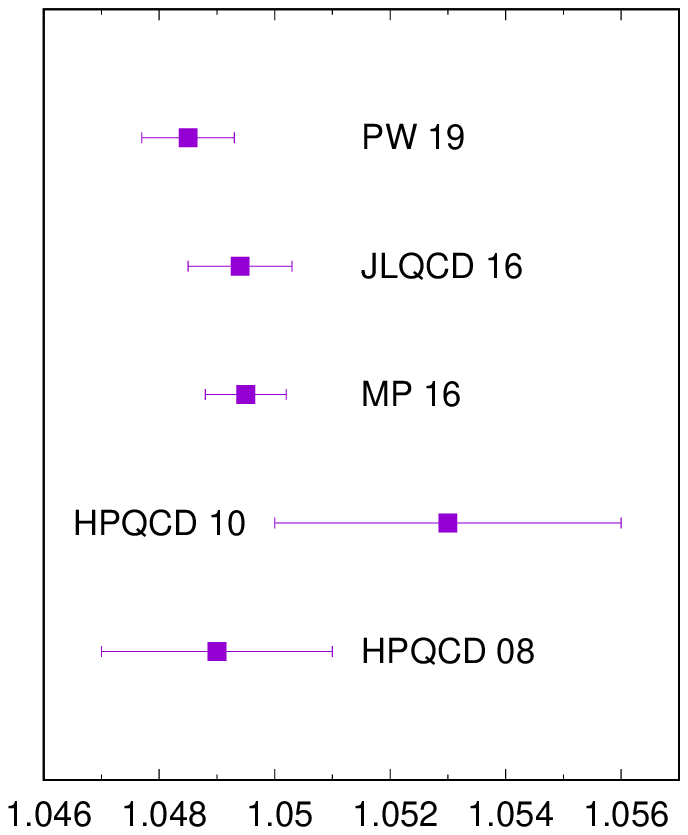}
\caption{Comparison of different lattice results for $R_4$ (left), $R_6/R_8$ (center) and $R_8/R_{10}$ (right).
Shown are the results of HPQCD collaborations from Refs. \cite{Allison:2008xk,McNeile:2010ji} labeled as HPQCD 08 and HPQCD 10,
as well as the results from JLQCD collaboration \cite{Nakayama:2016atf}.
The error on $R_6/R_8$ and $R_8/R_{10}$ for HPQCD 10 was obtained by propagating the errors on $R_6,~R_8$ and $R_{10}$ from Ref. \cite{McNeile:2010ji}.
}
\label{fig:comp}
\end{figure}

Before discussing the lattice results on $\alpha_s$ it is worthwhile to compare the different continuum
extrapolated lattice results for the reduced moments.
In Fig. \ref{fig:comp} I compare the continuum results for $R_4$, $R_6/R_8$ and $R_8/R_{10}$ from
different lattice studies for $m_h=m_c$. The new 3 flavor HISQ result \cite{Petreczky:2019ozv} (PW 19) agrees with HPQCD results, published in 2008 \cite{Allison:2008xk}
and 2010 \cite{McNeile:2010ji} and labeled as HPQCD 08 and HPQCD 10, 
but is higher than 
previous 3 flavor HISQ result from Ref. \cite{Maezawa:2016vgv}, denoted as MP 16. This is due to the fact that in Ref. \cite{Maezawa:2016vgv} simple
$a^2$ and $a^2+a^4$ continuum extrapolations have been used, which cannot capture the correct $a$-dependence of $R_4$ (see discussions in Ref. \cite{Petreczky:2019ozv}).
The ratio $R_6/R_8$ from the new 3 flavor HISQ calculation (PW 19) and the domain wall fermion calculation
JLQCD \cite{Nakayama:2016atf} (JLQCD 16) are smaller than the HPQCD results 
published in 2008 and 2010 (labeled as HPQCD 08 and HPQCD 10).
This could be due to the fact that in the analysis of HPQCD only few data points were available for small enough $a m_{h0}$.
For the same reason the MP 16 result for $R_6/R_8$ is larger (see discussions in Ref. \cite{Petreczky:2019ozv}).
Finally for $R_8/R_{10}$ all lattice results agree within errors, though there is some tension with HPQCD 10 results.
\begin{table}
\begin{tabular}{|l|lllll|}
\hline
$m_h$    & $R_4$              & $R_6/R_8$           &  $R_8/R_{10}$        & av.         & $\Lambda_{\overline{MS}}^{n_f=3}$ \\
\hline
$1.0$ & 0.3815(55)(30)(22) & 0.3837(25)(180)(40) &  0.3550(63)(140)(88) & 0.3788(65)  & 315(9)  \\
$1.5$ & 0.3119(28)(4)(4)   & 0.3073(42)(63)(7)   &  0.2954(75)(60)(17)  & 0.3099(48)  & 311(10) \\
$2.0$ & 0.2651(28)(7)(1)   & 0.2689(26)(35)(2)   &  0.2587(37)(34)(6)   & 0.2649(29)  & 285(8)  \\
$3.0$ & 0.2155(83)(3)(1)   & 0.2338(35)(19)(1)   &  0.2215(367)(17)(1)  & 0.2303(150) & 284(48) \\
\hline
\end{tabular}
\caption{The values of $\alpha_s(\mu=m_h)$ for different heavy quark masses, $m_h$ extracted 
from $R_4$, $R_6/R_8$ and $R_8/R_{10}$. The heavy quark mass is given in units of $m_c$.
The first, second and third errors correspond to the lattice error, the perturbative truncation error and the error
due to the gluon condensate.
In the fifth column the averaged value of $\alpha_s$ is shown (see text)
The last column gives the value of $\Lambda_{\overline{MS}}^{n_f=3}$ in MeV.}
\label{tab1}
\end{table}
Thus, there is consensus on the value of $R_4$, which is one of the quantities used to extract the strong coupling constant.
As we will see below $\alpha_s$ extracted from the ratios $R_6/R_8$ and $R_8/R_{10}$ have much larger errors.
Therefore, the differences seen in the values of the ratios are not the main reason for the differences
in the final $\alpha_s$ values.

\begin{figure}
\includegraphics[width=9cm]{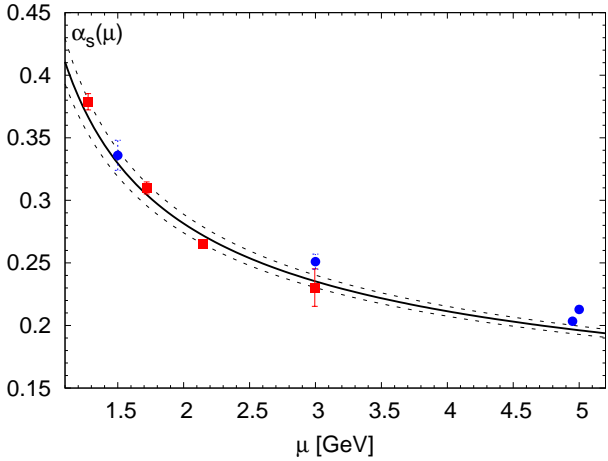}
\caption{The running coupling in three flavor QCD constant 
corresponding to $\Lambda_{\overline{MS}}^{n_f=3}=301(16)$ MeV.
The solid line corresponds to the central value, while the dashed lines correspond to the error band. The red
squares show the lattice results of this work. 
The blue circles from left to right correspond to the determination of $\alpha_s$
for the static quark anti-quark potential \cite{Bazavov:2014soa} and from the moments of quarkonium correlators 
\cite{Allison:2008xk,McNeile:2010ji,Chakraborty:2014aca}. 
The result of Ref. \cite{McNeile:2010ji} has been shifted horizontally for better visibility.
}
\label{fig:running}
\end{figure}

To obtain the value of the strong coupling constant from the fourth reduced moments or the ratio of the moments one needs
to specify the scale in the perturbative expansion. There is only one relevant physical scale present in the calculations of the moments
of the quarkonium correlators, the heavy quark mass $m_h$.  Therefore, the choice $\mu=m_h(\mu_m=m_h)$ is the natural one. 
This assertion is supported by the fact that the perturbative coefficients $r_{nj}$ are the smallest for this choice
of the renormalization scale. Another choice used by HPQCD is $\mu=\mu_m=3 m_h(\mu=3 m_h)$. There is no obvious reason
for the choice $\mu=\mu_m$ unless both of these scale are equal to $m_h$. In general $\mu$ and $\mu_m$ should be varied
independently, which leads to an increase of the perturbative uncertainty \cite{Dehnadi:2015fra}. If one adopts the choice
$\mu=m_h(m_h)$ the determination of $\alpha_s$ from the fourth moments or the above ratios of the moments reduces to
solving non-linear equations. The values of the strong coupling constant obtained from $R_4$, $R_6/R_8$ and $R_8/R_{10}$ 
using 3 flavor HISQ calculations \cite{Petreczky:2019ozv} at
different quark masses are presented in Table \ref{tab1}. 
As can been seen from the table the determination of the strong coupling constant from $R_6/R_8$
and $R_8/R_{10}$ has typically larger errors. Both the perturbative errors and the errors due to the gluon condensate
decrease with increasing $m_h$, as expected.
Some tension in the values of $\alpha_s$ determined
from different quantities and at different quark masses can be seen in Table \ref{tab1}. 
In particular, the central value of $\alpha_s$ determined from the ratio $R_8/R_{10}$ seems
to be consistently lower than the one determined from $R_4$ and $R_6/R_8$. Similar 
trend was observed in Refs. \cite{Allison:2008xk}.
The weighted average of
different determinations was calculated to obtain our final result at given $m_h$ \cite{Petreczky:2019ozv}. Having determined $\alpha(\mu=m_h)$ one can calculate
the value of $m_h$ itself from the higher order moments $R_n,~n \ge 6$. This is discussed in Ref. \cite{Petreczky:2019ozv}. Once $m_h$ is
determined we know the running coupling constant at low energies, which is shown in Fig. \ref{fig:running}. In the figure
I compare the running coupling constant determined this way (squares) with other lattice determinations (circles), including 
$\alpha_s$ from the static quark anti-quark potential \cite{Bazavov:2014soa} and HPQCD results from the moments of quarkonium
correlators \cite{Allison:2008xk,McNeile:2010ji,Chakraborty:2014aca}.
Furthermore, we can also determined the 3-flavor
$\Lambda$ parameter, $\Lambda_{\overline{MS}}^{n_f=3}$, which is in the last column of Table \ref{tab1}. As one can see
from the table the value of $\Lambda_{\overline{MS}}^{n_f=3}$ obtained at $m_h=2 m_c$ is significantly lower than 
the corresponding values obtained at $m_h=m_c$ and $m_h=1.5m_c$. Taking the weighted average of the $\Lambda$ parameters in
Table \ref{tab1} leads to the final result \cite{Petreczky:2019ozv}
\begin{equation}
\Lambda_{\overline{MS}}^{n_f=3}=301 \pm 16 ~ {\rm MeV}.
\end{equation}
\begin{figure}
\includegraphics[width=6cm]{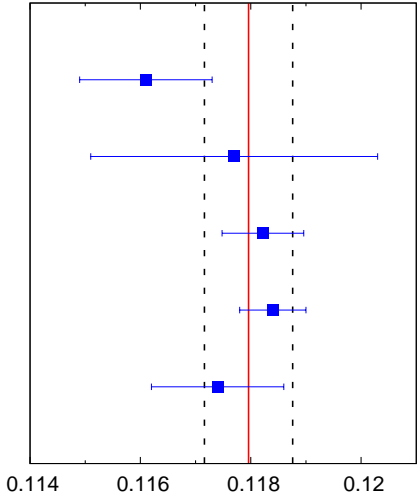}
\caption{The summary of $\alpha_s(M_Z,n_f=5)$ obtained from the lattice calculations of the moments
of quarkonium correlators. The vertical solid line represents the weighed average of different values, while
the vertical dashed lines correspond to the estimated uncertainty of $\alpha_s$ (see text).}
\label{fig:as_sum}
\end{figure}
The error in the above value was determined from the spread of the individual determinations around the average. 
From the value of $\Lambda_{\overline{MS}}^{n_f=3}$ one can determine the value of $\alpha_s$ in the 5 flavor theory at the
scale of the $Z$ boson mass $\mu=m_Z$ using the 4-loop evolution and the decoupling at the charm and bottom threshold as implemented
in the RunDeC package \cite{Chetyrkin:2000yt}. With this we get 
\begin{equation}
\alpha_s(M_Z,n_f=5)=0.1161(12).
\end{equation}
The above result is lower than the PDG value \cite{PDG18} and the FLAG value \cite{FLAG16}. It is also lower than $\alpha_s$ determined
from the moments of quarkonium correlators by HPQCD collaboration \cite{Allison:2008xk,McNeile:2010ji,Chakraborty:2014aca}. 
On the other hand it agrees with the result of JLab collaboration
\cite{Nakayama:2016atf} as well as the $\alpha_s$ determined from the energy of static quark anti-quark pair 
\cite{Bazavov:2012ka,Bazavov:2014soa,Takaura:2018vcy,Takaura:2018lpw}. In Fig. \ref{fig:as_sum} I summarize different $\alpha_s$
determinations using moments of quarkonium correlators. Averaging over different lattice results one gets $\alpha_s(M_z,n_f=5)=0.11796(40)$
with $\chi^2/df \simeq 0.8$. 
The error in this result may be too small since there are systematic errors common to all of the calculations. Perhaps doubling this error
gives a more realistic estimate of the uncertainty, which is indicated by vertical dashed lines in Fig. \ref{fig:as_sum}. 
Within this uncertainty all lattice results are in rough agreement.

\section{Conclusions}
In this proceeding contribution I reviewed the determination of the strong coupling constant
$\alpha_s$ from the moments of quarkonium correlators. The recent determination 
that is based on 3 flavor lattice QCD calculations with HISQ action was used as illustrative
example. The method has two challenges. One challenge is to obtain sufficiently precise
continuum extrapolation of the relevant moments or their ratio. The other challenge
is controlling the perturbative truncation errors and the scale dependence. 
The recent analysis gives smaller value of $\alpha_s$ than the previous lattice
determinations though with estimated errors it still agrees with the averaged value of
$\alpha_s$ from the moment method.

\bibliographystyle{JHEP}
\bibliography{pap_bib.bib}

\end{document}